\documentclass[]{spie}  
\usepackage[]{graphicx}
\usepackage[]{amssymb,amsmath}
\usepackage[]{caption}
\usepackage{xspace} 
\def\bicepp{{\sc Bicep2}\xspace}
\def\bicep{{\sc Bicep}\xspace}

\def\spider{{\sc Spider}\xspace}
\def\spud{{\sc Spud}\xspace}
\title{Optical Characterization of the Keck Array Polarimeter at the South Pole}

\author{A. G. Vieregg\supit{a}, P. A. R. Ade\supit{b}, R.~Aikin\supit{c}, C.~Bischoff\supit{a}, J.~J.~Bock\supit{c,d}, J.~A.~Bonetti\supit{d}, K.~J.~Bradford\supit{a}, J.~A.~Brevik\supit{c}, C.~D.~Dowell\supit{c,d}, L.~Duband\supit{e}, J.~P.~Filippini\supit{c}, S.~Fliescher\supit{f}, S.~R.~Golwala\supit{c}, M.~S.~Gordon\supit{a}, M.~Halpern\supit{g}, G.~Hilton\supit{h}, V.~V.~Hristov\supit{c}, K.~Irwin\supit{h}, S.~Kernasovskiy\supit{i}, J.~M.~Kovac\supit{a}, C.~L.~Kuo\supit{i}, E.~Leitch\supit{j}, M.~Lueker\supit{c}, T.~Montroy\supit{k}, C.~B.~Netterfield\supit{l}, H.~T.~Nguyen\supit{c,d}, R.~O'Brient\supit{c,d}, R.~W.~Ogburn~IV\supit{i}, C.~Pryke\supit{f}, J.~E.~Ruhl\supit{k}, M.~Runyan\supit{c}, R.~Schwarz\supit{f}, C.~Sheehy\supit{f,j}, Z.~Staniszewski\supit{c,d}, R.~Sudiwala\supit{b}, G.~Teply\supit{c}, J.~Tolan\supit{i}, A.~D.~Turner\supit{d}, P.~Wilson\supit{d}, and~C.~L.~Wong\supit{a}
\skiplinehalf
\supit{a}Harvard-Smithsonian Center for Astrophysics, 60 Garden Street, Cambridge, MA 02138, USA; \\
\supit{b}Dept. of Physics and Astronomy, University of Wales, Cardiff, CF24 3YB, Wales, UK; \\
\supit{c}California Institute of Technology, 1200 E. California Blvd., Pasadena, CA 91125, USA; \\
\supit{d}Jet Propulsion Laboratory, 4800 Oak Grove Dr., Pasadena, CA 91109, USA; \\
\supit{e}Service des Basses Tempratures, DRFMC, CEA-Grenoble, 17 rue des Martyrs, 38054~Grenoble Cedex 9, France; \\
\supit{f}School of Physics \& Astronomy, University of Minnesota, 116 Church Street S.E., Minneapolis, MN 55455, USA; \\
\supit{g}Department of Physics \& Astronomy, University of British Columbia, 6224~Agricultural~Road, Vancouver, BC V6T1Z1, Canada; \\
\supit{h}NIST Quantum Devices Group, 325 Broadway, Boulder, CO 80305, USA; \\
\supit{i}Stanford University, 382 Via Pueblo Mall, Stanford, CA 94305, USA; \\
\supit{j}University of Chicago, KICP, 933 E. 56th St., Chicago, IL 60637, USA; \\
\supit{k}Physics Department, Case Western Reserve University, Cleveland, OH 44106, USA; \\
\supit{l}Department of Physics, University of Toronto, Toronto, ON M5S 1A7, Canada; \\
}

\authorinfo{Send correspondence to A. G. Vieregg, 60 Garden Street, MS 42, Cambridge, MA 02138 USA, E-mail: avieregg@cfa.harvard.edu}

\begin{document}
  \maketitle

\begin{abstract}
The Keck Array (\spud) is a set of microwave 
polarimeters that observes from the South Pole at degree angular scales in search of a signature of 
Inflation imprinted as 
B-mode polarization in the Cosmic Microwave Background (CMB).  The first three Keck Array receivers were deployed
during the 2010-2011 Austral summer, followed by two new receivers in the 2011-2012 summer season, completing the full
five-receiver array.  All five receivers are currently observing at 150~GHz.  
The Keck Array employs the field-proven 
\bicep/\bicepp strategy of using small, cold, on-axis refractive optics, providing excellent control 
of systematics while maintaining a large field of view.  This design allows for full characterization of far-field
optical performance using microwave sources on the ground.  We describe our efforts to characterize the main 
beam shape and beam shape mismatch between co-located 
orthogonally-polarized detector pairs, and discuss the implications of measured
differential beam parameters on temperature to polarization leakage in CMB analysis.

\end{abstract}




\section{INTRODUCTION}
\label{sec:intro}
Cosmological Inflation is a theory that describes the entire observable Universe as a microscopic
volume that underwent violent, exponential expansion during the first
fraction of a second.  Inflation is supported by the flatness and extreme uniformity of the Universe
observed through measurements of the Cosmic Microwave Background (CMB)~\cite{boomerang,wmapFlat}.  
Measurements of the polarization of the CMB could prove to be an impressive tool for 
probing the epoch of Inflation. 
A generic prediction of Inflation is the production of a Cosmic Gravitational-Wave Background, 
which in turn would imprint a faint but unique 
signature in the polarization of the CMB that has a curl component~\cite{kamionkowski, seljak}.  This
curl component of the polarization field is commonly called B-mode polarization, while the curl-free
component, dominated by production due to density fluctuations at the time of last 
scattering, is called E-mode polarization.  
The strength of the B-mode polarization signature depends on the energy scale of Inflation,
and would be detectable 
if Inflation occurred near the energy scale at which the fundamental forces unify ($\sim 10^{16}$~GeV).

The Keck Array, also called \spud, is a set of five degree-scale 
microwave polarimeters that is currently observing the CMB from the Martin A. Pomerantz Observatory at the South 
Pole in search of a B-mode polarization signature from Inflation~\cite{sheehy}.
Each of the five receivers has 512~Transition Edge Sensor (TES) detectors, 16 of which are dark, leaving 
496~detectors that are coupled to planar arrays of slot 
antennas, for a total of 2480~optical detectors in the entire instrument.   Receivers utilize a compact, 
on-axis refracting telescope design.

The first three Keck Array receivers
were deployed during the 2010-2011 Austral summer and two new receivers followed in the 2011-2012 deployment
season.  The first season of observation with the full five-receiver array is currently underway,
with all receivers observing at 150~GHz.  The modular design of the Keck Array 
allows for future upgrades to include replacement of individual receivers to provide 
additional frequency coverage at 100~GHz and 220~GHz.
The Keck Array leverages field-proven techniques employed for the \bicep and \bicepp telescopes,
but with a vastly increased number of detectors, 
leading to increased sensitivity to the tiny Inflationary B-mode signal.
The current upper limit on the B-mode amplitude in the CMB is set by the Keck Array's predecessor 
experiment, \bicep~\cite{chiang, takahashi}, and corresponds to $r < 0.72$ at 95\% confidence level, 
where $r$ is the tensor-to-scalar ratio.
The Keck Array aims to reach a sensitivity corresponding to $r=0.01$, where gravitational lensing of E-modes 
into B-modes should begin to be comparable in strength to the Inflationary signal.

As sensitivity dramatically improves with each generation of experiments, control 
of systematics becomes increasingly important.  Precise characterization of the optical performance of
Keck Array receivers is critical to reach these ambitious sensitivity goals.  In the simplest mapmaking schemes, 
differential beam 
effects between co-located orthogonally-polarized pairs of detectors
can lead to leakage of the CMB temperature signal into the much smaller B-mode signal, 
potentially limiting the ability of the Keck Array to reach its design sensitivity if these systematics
are not well understood.  

Characterizing the beam pattern of each of the 2480~Keck Array detectors in the far field presents a challenge 
in both data acquisition and reduction. We describe here our effort to characterize the Keck Array optical performance 
through an extensive ground-based precision beam mapping campaign at the South Pole,
and discuss our understanding of the cause of measured beam non-idealities and 
our strategy for mitigating the effect of measured differential beam components in CMB analysis.

Four companion papers are also presented at this conference, focusing on the status of \bicepp 
and the Keck Array (Ogburn {\it et al.}~\cite{ogburnSPIE}), the sensitivity of the Keck Array 
(Kernasovskiy {\it et al.}~\cite{kernasovskiySPIE}),
the performance of the dual-polarization planar antenna array (O'Brient {\it et al.}~\cite{obrientSPIE}),
and the thermal stability of \bicepp (Kaufman~{\it et~al.}\cite{kaufmanSPIE}).

\section{OPTICAL DESIGN}
\label{sec:optics}
\begin{figure}[ht]
   \begin{center}
   \begin{tabular}{c}
   \includegraphics[height=6cm]{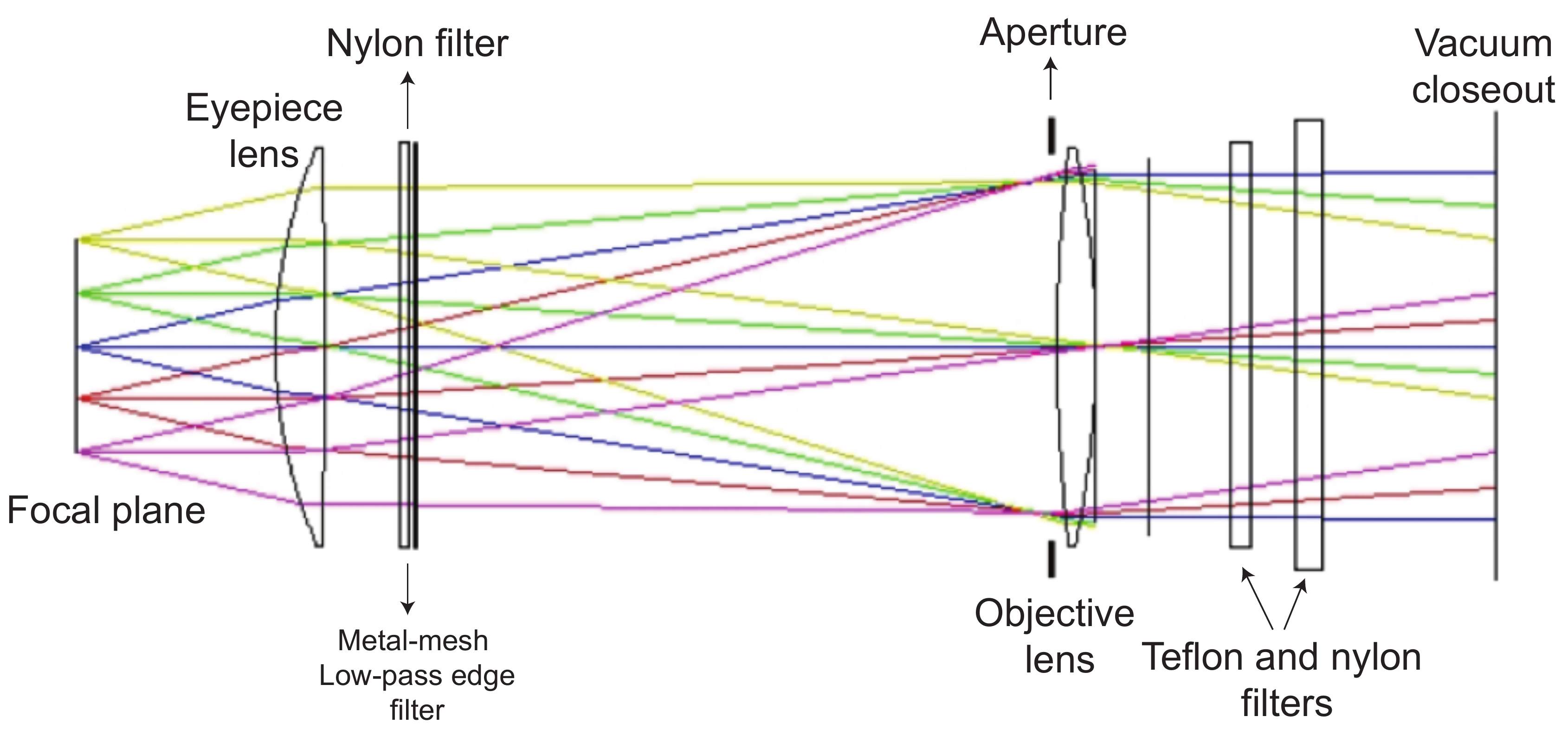}
   \end{tabular}
   \end{center}
   \caption{ \label{fig:opticsChain}
   A schematic of the \bicepp/Keck Array optical chain, from Aikin {\it et al.}~\cite{randol}. } 
   \end{figure}

Each Keck Array receiver is a compact, single-frequency, on-axis refractive telescope
with an aperture of 26.4~cm, designed to maintain tight control
of systematics. All optical elements are cold (4~K or 50~K) to maintain low and stable optical loading on the focal
plane.
Figure~\ref{fig:opticsChain} shows a schematic of the \bicepp/Keck Array optical chain.  
An in-depth discussion of the design of \bicepp optical elements can be
found in Aikin {\it et al.}~\cite{randol}.

The eyepiece and objective lenses are made of high-density polyethylene (HDPE)
and are designed to provide even illumination of the aperture, which is coincident with 
the objective lens.  The illumination at the edge of the aperture compared to the illumination
at the center of the aperture is designed to be -12.4~dB for 
detectors at the center of the focal plane.  Both lenses 
as well as the telescope housing and aperture are cooled to 4~K.
Two Teflon, two nylon, and one metal-mesh filter block IR radiation 
from reaching the focal plane.  The two Teflon filters
and one nylon filter are located in front of the objective lens, and are held at 50~K.  Another nylon filter and 
the metal-mesh filter sit at 4~K between
the objective and eyepiece lenses inside the telescope itself to further reduce loading.  Teflon has excellent
in-band transmission at cryogenic temperatures.  While nylon has higher in-band transmission loss, it has a steeper
transmission rolloff, providing significant reduction of far-IR loading on the sub-Kelvin stages.  The metal-mesh 
filter is a low pass filter with a cutoff at 250~GHz, providing additional blocking of out-of-band power.
 
Keck Array optical elements are anti-reflection coated with porous
Teflon with an index of refraction matched to the optical element 
and thickness matched to the observing frequency of each receiver, currently 150~GHz.  
Future upgrades to the array include receivers
that observe at 100~and 220~GHz, which require different anti-reflection coatings but
contain an otherwise identical optical system.

The vacuum window has a 32~cm clear aperture, making design and construction of strong and durable
vacuum windows a challenge.
Keck Array vacuum windows are made of Zotefoam HD30, a nitrogen-expanded polyethylene foam that was
chosen for its high microwave transmission, its strength against deflection under vacuum,
and its adhesion strength to epoxy used to bond the foam to an aluminum frame. 

To reduce sidelobe pickup, individual co-moving ground shields are installed in front of each receiver's vacuum window.  
These forebaffles are coated on the inside with HR10 microwave absorber and a weatherproofing foam, providing good 
termination of sidelobes.  The forebaffles intersect radiation at $9.5^{\circ}$ off of the boresight axis 
from the edge of the vacuum window.

The \bicepp and Keck Array optical designs are identical except for a few small differences. 
The material used for the vacuum window for \bicepp is Zotefoam PPA30, but is 
Zotefoam HD30 for the Keck Array. \bicepp has a larger forebaffle than Keck Array receivers.
The exact configuration of the IR blocking filters is different between the two experiments;
\bicepp has one Teflon filter at 40~K, another at 100~K, and only has one nylon filter (at 4~K).


\section{OPTICAL CHARACTERIZATION}
\label{sec:beams}
\subsection{Near-Field Beam Characterization}
\begin{figure}[ht]
   \begin{center}
   \begin{tabular}{c}
   \includegraphics[height=5cm]{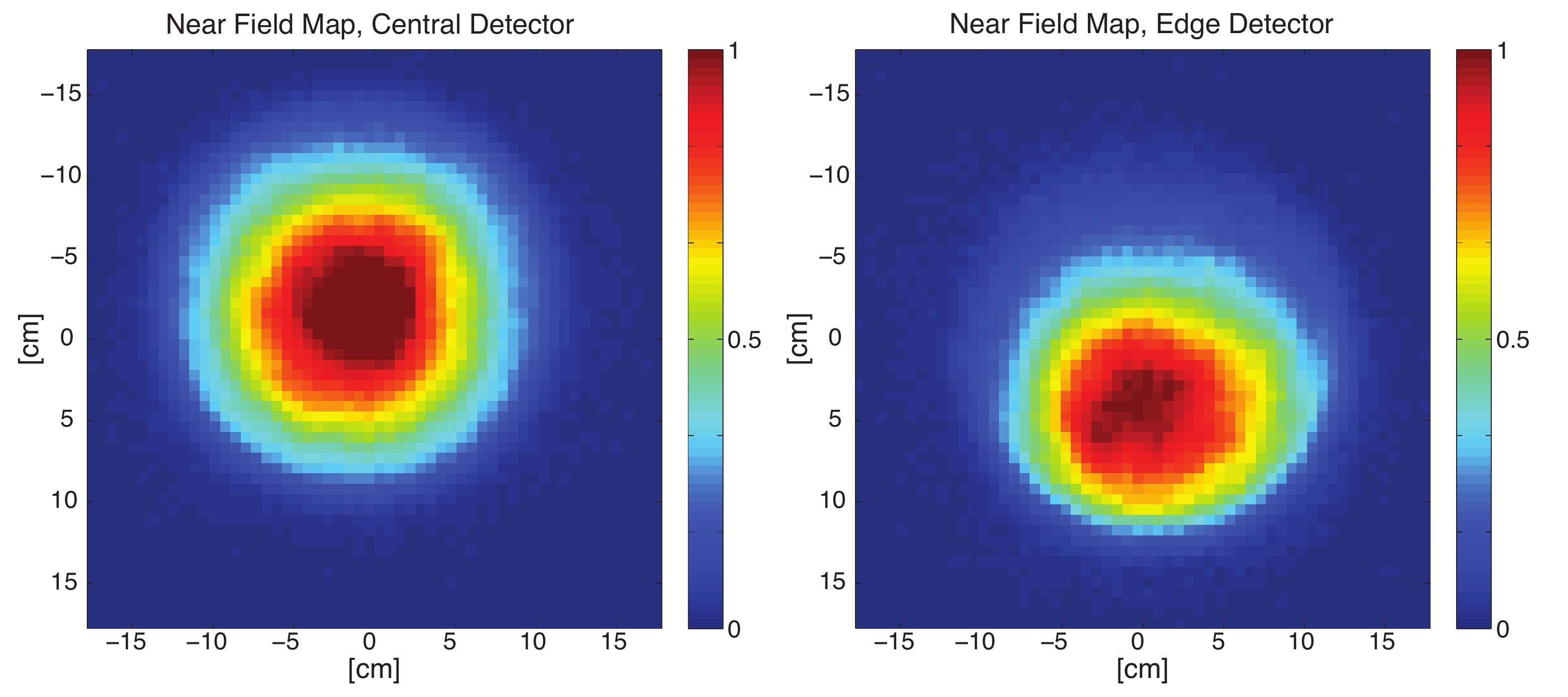}
   \end{tabular}
   \end{center}
   \caption{ \label{fig:nearField}
Measurement of the near-field beam pattern for two example Keck Array channels, measured above the 
aperture stop at the vacuum window.  Left: A detector near the center of the
focal plane; Right: A detector near the edge of the
focal plane,
showing significant truncation by the aperture (worst case).}
   \end{figure}

To characterize aperture illumination, we measure the near-field beam pattern of each Keck Array 
detector using a chopped thermal source
mounted on an x-y translation stage attached to the cryostat above the vacuum window, as close to
the aperture stop of the telescope as possible.  In practice, the 
source is about 30~cm above the aperture.  Figure~\ref{fig:nearField} shows the beam pattern of two example detectors
in the near field.  The left panel shows the beam pattern of a 
detector near the center of the focal plane, which evenly 
illuminates the aperture. The right panel shows the beam pattern of a detector 
near the edge of the focal 
plane that is significantly truncated by the aperture because of non-ideal 
beam pointing at the focal plane (worst case).
This type of truncation translates to some ellipticity in
the beam pattern in the far field and only affects a small fraction of detectors.

\subsection{Polarization Angle and Polarization Efficiency Characterization}
We must know the polarization angle, $\psi$, and cross-polar response, $\epsilon$, of each detector
to correctly construct polarization maps of the sky.  The polarization efficiency is calculated from
the cross-polar response as \mbox{$(1-\epsilon)/(1+\epsilon)$}.  \bicep and \bicepp measured these polarization
parameters using a polarized broadband amplified microwave source 
by rotating the telescope about its boresight while the source polarization remained fixed~\cite{takahashi}.
For \bicepp, the polarization efficiency was measured to be $< 1$\%, with leakage dominated by inductive 
crosstalk between front-end SQUIDs in the readout system~\cite{randol}.
This technique, however, would be tedious with the Keck Array because as we rotate the drum that houses the 
receivers to rotate the receivers
about their boresights, the physical location of each receiver also moves, making the data set much more difficult
to take and analyze than it was for \bicep and \bicepp.

\begin{figure}[h]
   \begin{center}
   \begin{tabular}{c}
   \includegraphics[height=5cm]{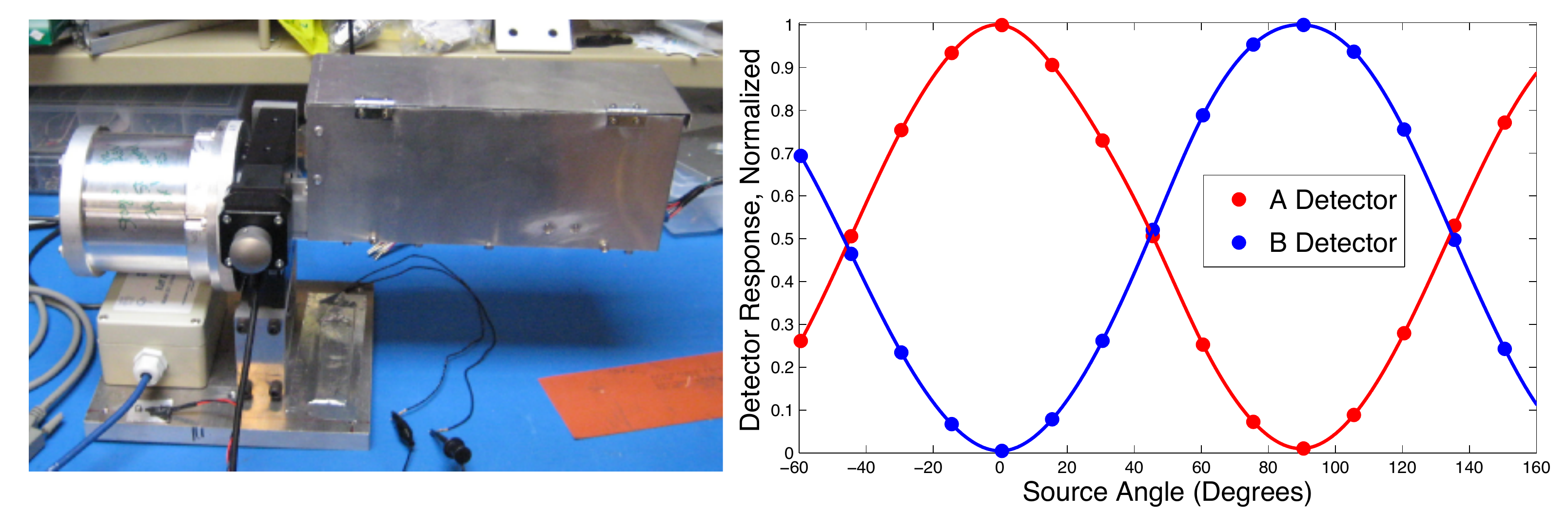}
   \end{tabular}
   \end{center}
   \caption{ \label{fig:rps}
     Left: The rotating polarized amplified thermal broadband noise source used for polarization characterization.  
     Right: Polarization modulation
     vs. source angle of an example detector 
     pair from \bicepp, measured using the rotating polarized source.
}
   \end{figure}

To measure the polarization angle and cross-polar response of each detector without having
to rotate the drum, we have developed a new rotating
polarized broadband amplified microwave source (see Figure~\ref{fig:rps}).
The source emits radiation in the 140-160~GHz range, 
designed to cover the passband of the current Keck Array receivers.  
A $50~\Omega$ load provides room-temperature
thermal noise at the input of the first stage of amplification (80~dB).  A series of frequency multipliers, 
amplifiers, and filters bring the output frequency to the desired range (140-160~GHz). Linearly polarized 
radiation is emitted by a 15~dB gain horn antenna, and is further polarized by a free-standing wire grid, 
yielding cross-polar leakage of the source $< 0.03$\%.
Two variable attenuators in series 
allow for control of output power over a large dynamic range, making the source useful for far-field beam mapping
as well as sidelobe mapping with the source closer to the receiver.  A microwave switch chops the source 
at $\sim10$~Hz. 
The entire source is mounted on a stepped rotating stage and 
has a total positional repeatability~$< 0.01^{\circ}$.

We used the rotating polarized source to repeat a measurement of detector polarization angles and cross-polar 
response for \bicepp
in February 2012.  An example of the polarization modulation vs. source angle for one pair of \bicepp
detectors is shown
in Figure~\ref{fig:rps}.  The orthogonally-polarized detectors in the pair
are $90^{\circ}$ out of phase.  We plan to make similar measurements
with the Keck Array during the upcoming 2012-2013~summer season.

\subsection{Far-Field Beam Characterization}
\subsubsection{Data Acquisition and Reduction}
\begin{figure}[ht]
   \begin{center}
   \begin{tabular}{c}
   \includegraphics[height=6cm]{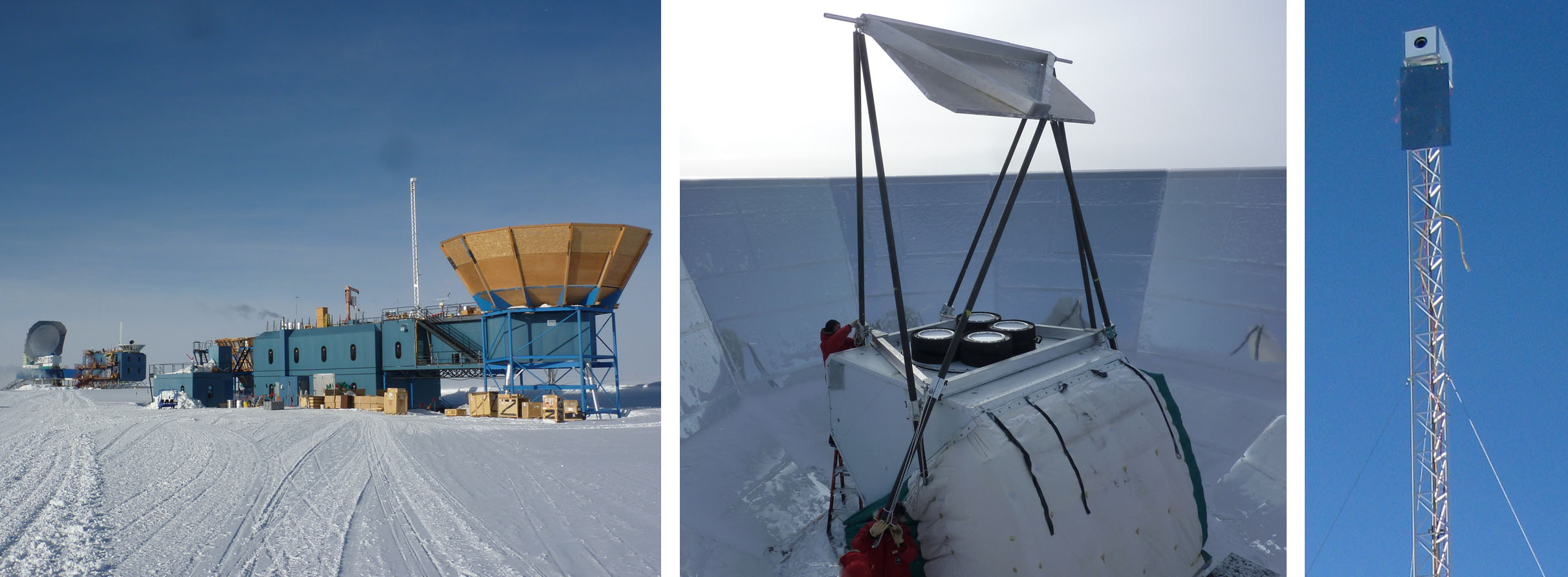}
   \end{tabular}
   \end{center}
   \caption{ \label{fig:darkSector}
The setup for measuring far-field beam pattern of Keck Array detectors 
in situ at the South Pole.  A chopped, broadband, 
microwave source broadcasts from a mast on the Dark Sector Laboratory (DSL), and a large aluminum honeycomb mirror 
is installed 
to redirect the beams of the Keck Array to the source.  
Left: The Keck Array in the foreground, with DSL in the background. 
Middle: The aluminum honeycomb mirror, installed on the Keck Array for beam measurements.  Right: The microwave source
mounted on a mast on the roof of DSL.}
   \end{figure}

Measuring the beam pattern of each of the 2480~detectors of the Keck Array in the far field 
presents a challenge and requires an extensive beam mapping campaign at the South Pole.  
Figure~\ref{fig:darkSector}
shows the setup used for measuring the beam pattern in the far field.  With all five 
receivers installed in the drum, we install a $1.2\times1.8$~m aluminum honeycomb mirror, 
flat to 0.2~mm across the mirror, 
mounted on carbon-fiber rods.  The mirror assembly was designed to be installed with no overhead crane;
the entire system weighs less than 150~kg.
The mirror redirects the beams over the top of the ground shield and 
to a chopped thermal microwave source with an aperture of 20~cm mounted on a 10~m tall mast on the Dark Sector 
Laboratory, 211~m away.  The
thermal source chops between a flat mirror directed to zenith ($\sim 15$~K) and 
ambient ($\sim 260$~K) at a tunable frequency, set to be 10~Hz.

The size of the mirror allows us to map only three receivers at a time.  By rotating the drum below the mirror, we 
can move selected receivers into a position under the mirror for beam mapping.  For each receiver, we 
take data at five different drum angles so that we can check that a rotation of a receiver underneath the mirror
does not affect the measurement.  The five drum rotations used for each receiver
are separated by 36$^{\circ}$, for a total drum rotation coverage of 144$^{\circ}$ for each receiver, 
the maximum we could achieve
using any single mirror position.

The data set discussed here 
consists of five measurements (one at each of five drum rotations) of each of the five Keck Array receivers,
each with 496 optically-coupled detectors.  This data set was taken in February 2012, at the end of the most recent 
summer season. We filter the chop reference signal to match the filtering that occurs in the readout system and then 
demodulate the timestream data.  The reflection off the mirror and parallax effects are handled with a pointing model
that describes the Keck Array mount system as well as the mirror used for beam mapping. 

\subsubsection{Beam Parameters}
\begin{figure}[h]
   \begin{center}
   \begin{tabular}{c}
   \includegraphics[height=16cm]{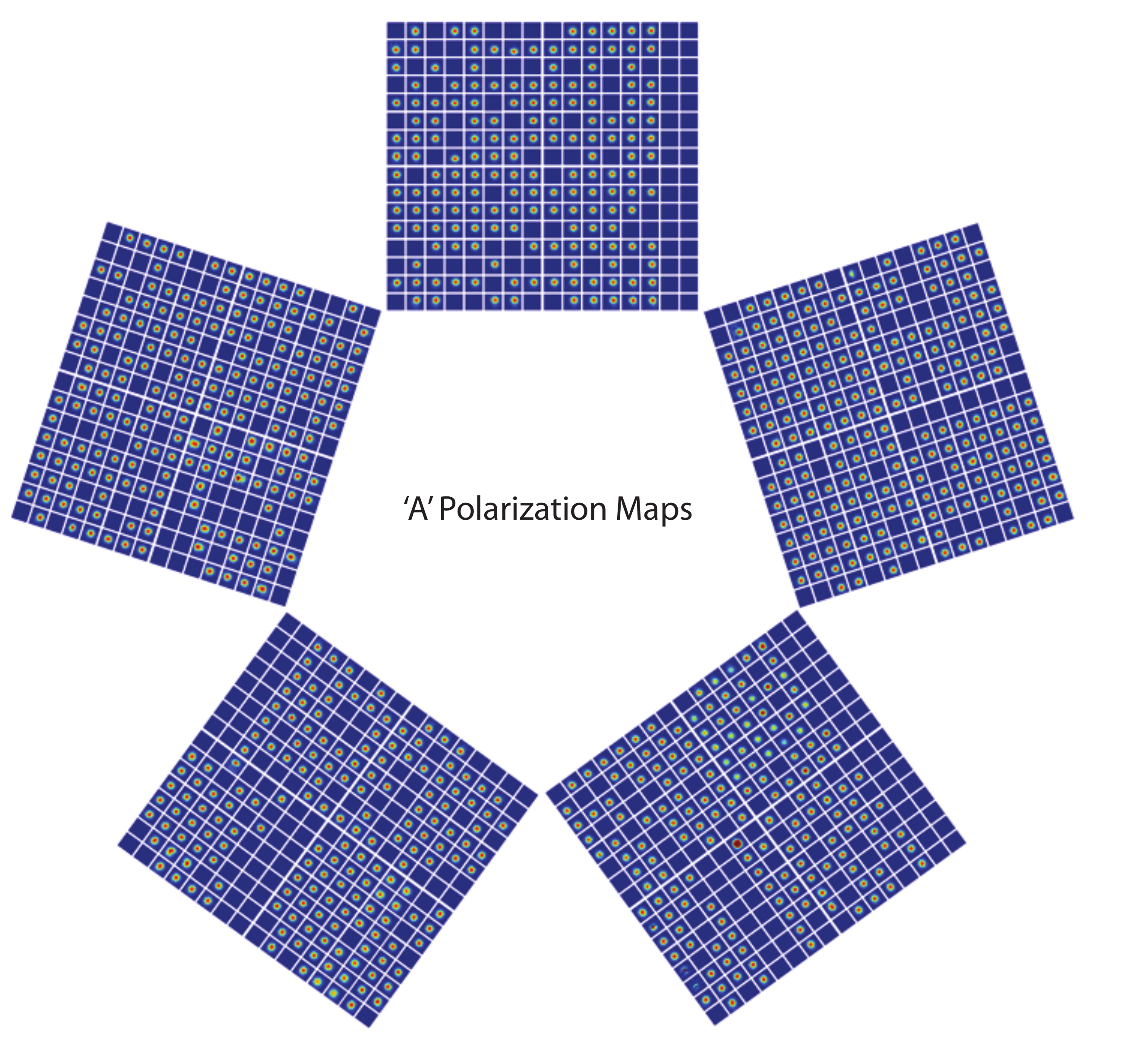}
   \end{tabular}
   \end{center}
   \caption{ \label{fig:5rxMap}
     Measured beam maps of A polarization detectors for the entire array.  The maps
     are arranged in a layout that represents the location and orientation of the five receivers in the drum, and 
     does not represent the field of view.  In the far field, the five receivers have completely overlapping 
     fields of view, and the beam centers
     are more closely spaced than shown here.
}
   \end{figure}

Figure~\ref{fig:5rxMap} displays measured beam maps for all detectors of 
a single polarization (denoted ``A'' polarization) over the entire array.
We fit an elliptical Gaussian to the main beam for each detector, according to
\begin{eqnarray}\label{eqn:gaussian}
g e^{-\frac{1}{2}(\vec{x}-\vec{r})\Sigma^{-1}(\vec{x}-\vec{r})}
\end{eqnarray}
where $\vec{r}$ is the location of the beam center, $g$ is the amplitude, and $\Sigma$ is the covariance matrix. 
Ellipticity parameters, $p$ and $c$, corresponding to amplitudes of the 
the ``plus'' and ``cross'' ellipticity orientations, 
are defined below in Equation~\ref{eqn:sigma}
as part of the covariance matrix.
\begin{eqnarray}\label{eqn:sigma}
\Sigma = \left[ \begin{array}{cc}
\sigma^2(1+p) & c\sigma^2 \\
c\sigma^2 & \sigma^2(1-p) 
\end{array}\right]
\end{eqnarray}

\begin{figure}[ht]
   \begin{center}
   \begin{tabular}{c}
   \includegraphics[height=5cm]{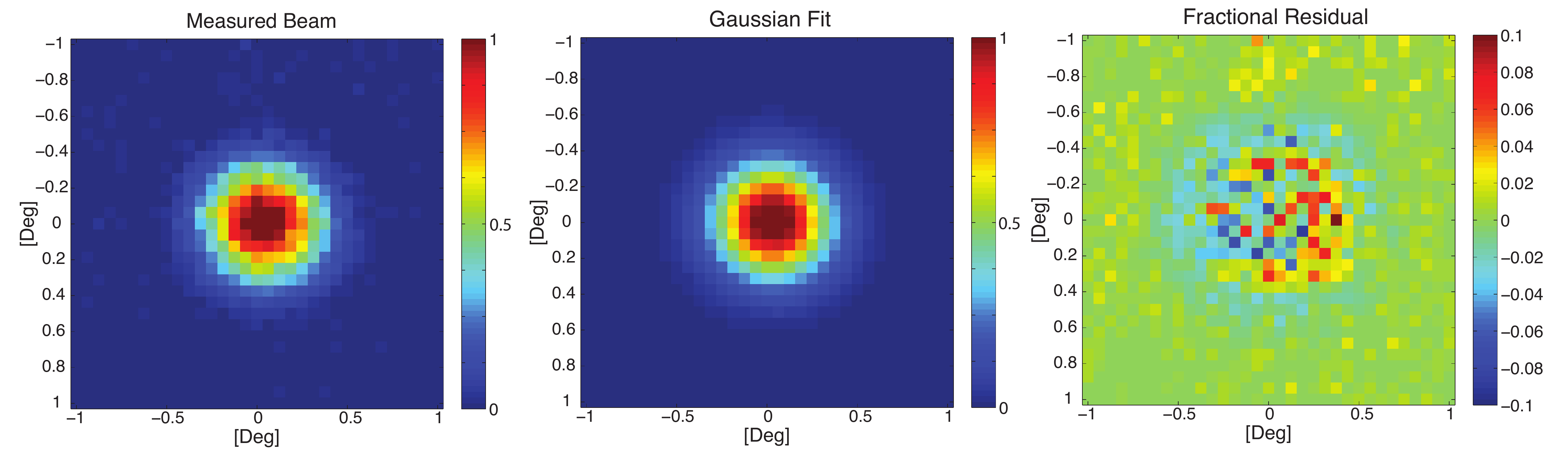}
   \end{tabular}
   \end{center}
   \caption{ \label{fig:farFieldBeam}
Left: Example measured far-field beam pattern, linear scale. Middle: Gaussian fit to measured beam pattern. 
Right: Fractional 
residual after subtracting the Gaussian fit in the middle panel.  Note: Right-hand panel has a different 
color scale than the left two panels.
}
   \end{figure}

\begin{figure}[ht]
   \begin{center}
   \begin{tabular}{c}
   \includegraphics[height=5cm]{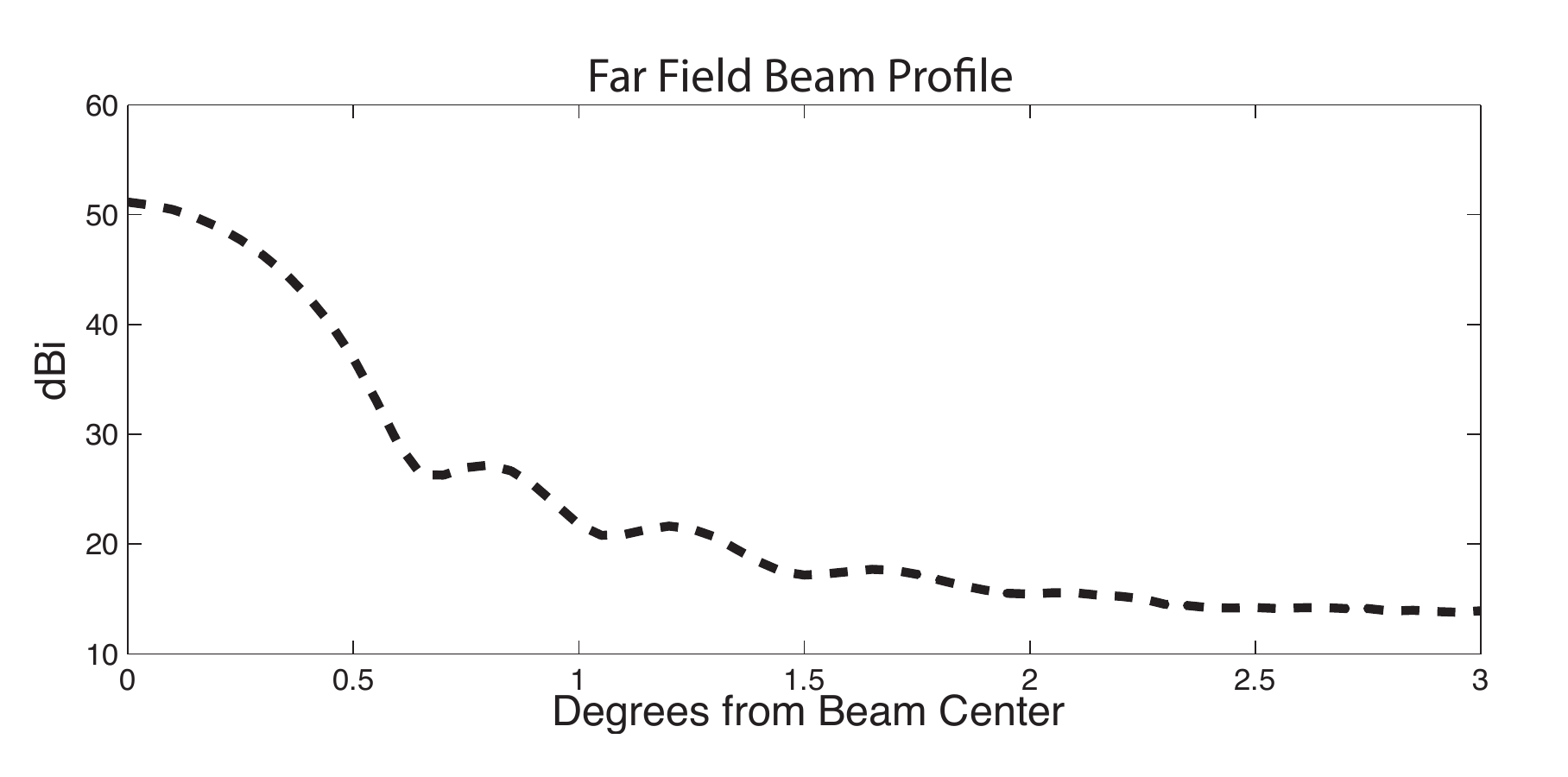}
   \end{tabular}
   \end{center}
   \caption{ \label{fig:farFieldBeamProfile}
Radial profile of the Keck Array far-field beam pattern, averaged over one receiver.  The profile shows the 
first few sidelobes and nulls before the beam pattern falls to the noise floor of the measurement.
}
   \end{figure}

Figure~\ref{fig:farFieldBeam} shows an example map,
the elliptical Gaussian fit, and the fractional residual after subtracting the fit.  
Average beam widths and ellipticities are
shown in Table~\ref{table:paramNumbers} for each receiver.   The average beam
width ($\sigma$) over the array is $0.215\pm0.007^{\circ}$.  Figure~\ref{fig:farFieldBeamProfile}
shows the measured far-field beam profile, averaged over all detectors in one Keck Array receiver.


\begin{table*}[h]
\small
\begin{center}
\begin{tabular}[c]{|l|c|c|c|c|c|}
\hline
 & \multicolumn{5}{|c|}{Keck Array 2012 Receiver-Averaged Beam Parameter Values}  \\
Parameter     & Receiver 0 & Receiver 1 & Receiver 2 & Receiver 3 & Receiver 4   \\
\hline\hline
Beam width ($\sigma$, degrees) & $0.214\pm0.005$ & $0.213\pm0.006$ & $0.213\pm0.006$ & $0.216\pm0.008$ & $0.218\pm0.013$ \\
Beam ellipticity$^1$  & $0.010\pm0.007$ & $0.012\pm0.006$ & $0.012\pm0.007$ & $0.013\pm0.010$ & $0.013\pm0.010$ \\
\hline
\multicolumn{6}{l}
{\scriptsize $^1$Beam ellipticity is defined as $(\sigma_{\mathrm{major}}-\sigma_{\mathrm{minor}})/(\sigma_{\mathrm{major}}+\sigma_{\mathrm{minor}})$, where $\sigma_{\mathrm{major}}$ and $\sigma_{\mathrm{minor}}$ are the eigenvalues of $\Sigma$.} \\
\end{tabular}
\end{center}
\caption{Receiver-averaged single beam parameters for all five Keck Array receivers with the standard 
deviation of these parameters across each receiver.  The spread is dominated by real detector-to-detector
scatter, not measurement uncertainty for individual detectors.}
\label{table:paramNumbers}
\end{table*} 
\subsubsection{Differential Beam Parameters}
\label{sec:parameters}
\begin{figure}[ht]
   \begin{center}
   \begin{tabular}{c}
   \includegraphics[height=5.5cm]{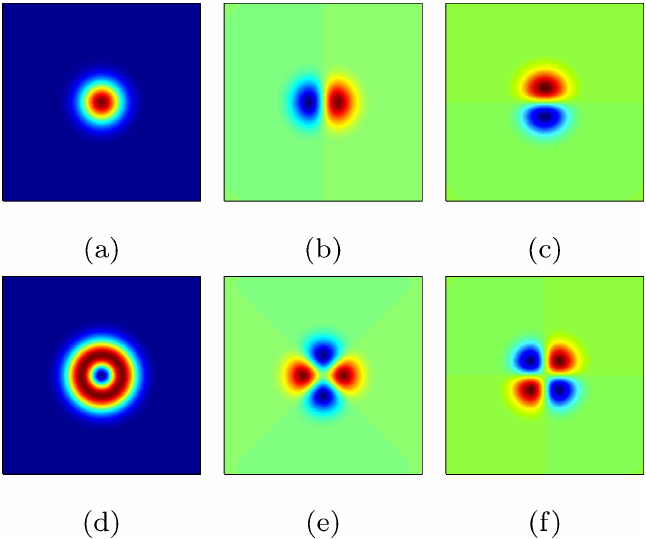}
   \end{tabular}
   \end{center}
   \caption{ \label{fig:6params}
Differential beam templates resulting in mismatch in (a) responsivity 
(b) x-position (c) y-position (d) beam width (e) 
ellipticity in plus (f) ellipticity in cross.  In the limit of small differential parameters, a differenced beam 
pattern constructed from the difference of two elliptical Gaussians
can be represented as a linear combination of each of these templates.  From Aikin {\it et al.}~\cite{methods}.
}
   \end{figure}

The beam pattern of a single detector can also be characterized as a set of 
perturbations
on an idealized circular Gaussian fit, with a nominal width ($\sigma_{\mathrm{n}}$) 
equal to the receiver-averaged value and a nominal beam 
center equal to the calculated center for the pair of detectors from the initial elliptical Gaussian fit.
We consider the first six perturbations, corresponding
to the templates shown in Figure~\ref{fig:6params}.  These six templates correspond to relative responsivity, 
x-position offset,
y-position offset, beam width, ellipticity in the ``plus'' orientation, and ellipticity in the ``cross'' orientation.

We calculate the regression coefficient for each measured individual beam pattern against each template map.
Differential beam parameters for a pair of co-located orthogonally-polarized 
detectors, which we denote ``A'' and ``B'' detectors, 
are then the difference of the regression coefficients for each detector.  
Table~\ref{table:parameters} gives the correspondence between these differential beam parameters 
and the associated elliptical Gaussian fit parameters for each beam map (Equations~\ref{eqn:gaussian} and~\ref{eqn:sigma}) 
in the limit of small perturbations. 

In this analysis, we extract differential beam parameters 
by calculating regression coefficients of individual detector beam maps against each of the template maps and then
differencing regression coefficients for each detector pair.
This 
proves convenient when predicting and mitigating beam mismatch-induced polarization, as
the leakage scales linearly with these differential parameters. 
A full description of this parameterization and its implementation in analysis can be 
found in Aikin~{\it et al.}~\cite{methods}.


\begin{table*}[h]
\small
\begin{center}
\begin{tabular}[c]{|l|c|}
\hline
Parameter               & Definition     \\
\hline\hline
Differential relative responsivity ($\delta g$)  &  $\Delta(g_A/g_{\mathrm{n}})/(g_B/g_{\mathrm{n}})$   \\
Differential pointing in $x$ ($\delta r_x$) & $ (\vec{r}_a-\vec{r}_b)\cdot \hat{x} /2\sigma_{\mathrm{n}} $\\
Differential pointing in $y$ ($\delta r_y$)  & $ (\vec{r}_a-\vec{r}_b)\cdot \hat{y} /2\sigma_{\mathrm{n}} $  \\
Differential beam width ($\delta \sigma $)  & $ (\sigma_A^2 - \sigma_B^2)/\sigma_{\mathrm{n}}^2$  \\
Differential ellipticity, plus orientation  ($\delta p$)   & $  (p_A - p_B)/2 $    \\
Differential ellipticity, cross orientation ($\delta c$) & $ (c_A - c_B)/2$  \\
\hline
\end{tabular}
\end{center}
\caption{The correspondence between differential beam parameters and the associated elliptical Gaussian fit parameters 
  for each beam map in the limit of small perturbations.  The two unitless parameters corresponding to 
  differential ellipticity, 
  $p$ and $c$, are the ellipticity parameters defined in Equation \ref{eqn:sigma}. 
  $\sigma_{\mathrm{n}}$ is the 
  receiver-averaged beam width, $\vec{r}_a$ and $\vec{r}_b$ are the vector from the averaged detector-pair 
  center to the A or B beam center, $\sigma_A$ and 
  $\sigma_B$ are the beam width for A and B, $g_A$ and $g_B$ are the responsivity for A and B, and $g_{\mathrm{n}}$ is the 
  nominal receiver-averaged responsivity.  This table is from Aikin {\it et al.}~\cite{methods}.}
\label{table:parameters}
\end{table*} 

We can calculate the regression coefficient for five of the templates shown in Figure~\ref{fig:6params} from beam maps, 
and extract differential beam parameters for each detector pair. The sixth coefficient, for the relative 
responsivity, is equivalent to differential gain which is calibrated frequently during CMB observations.

\begin{figure}[ht]
   \begin{center}
   \begin{tabular}{c}
   \includegraphics[height=9cm]{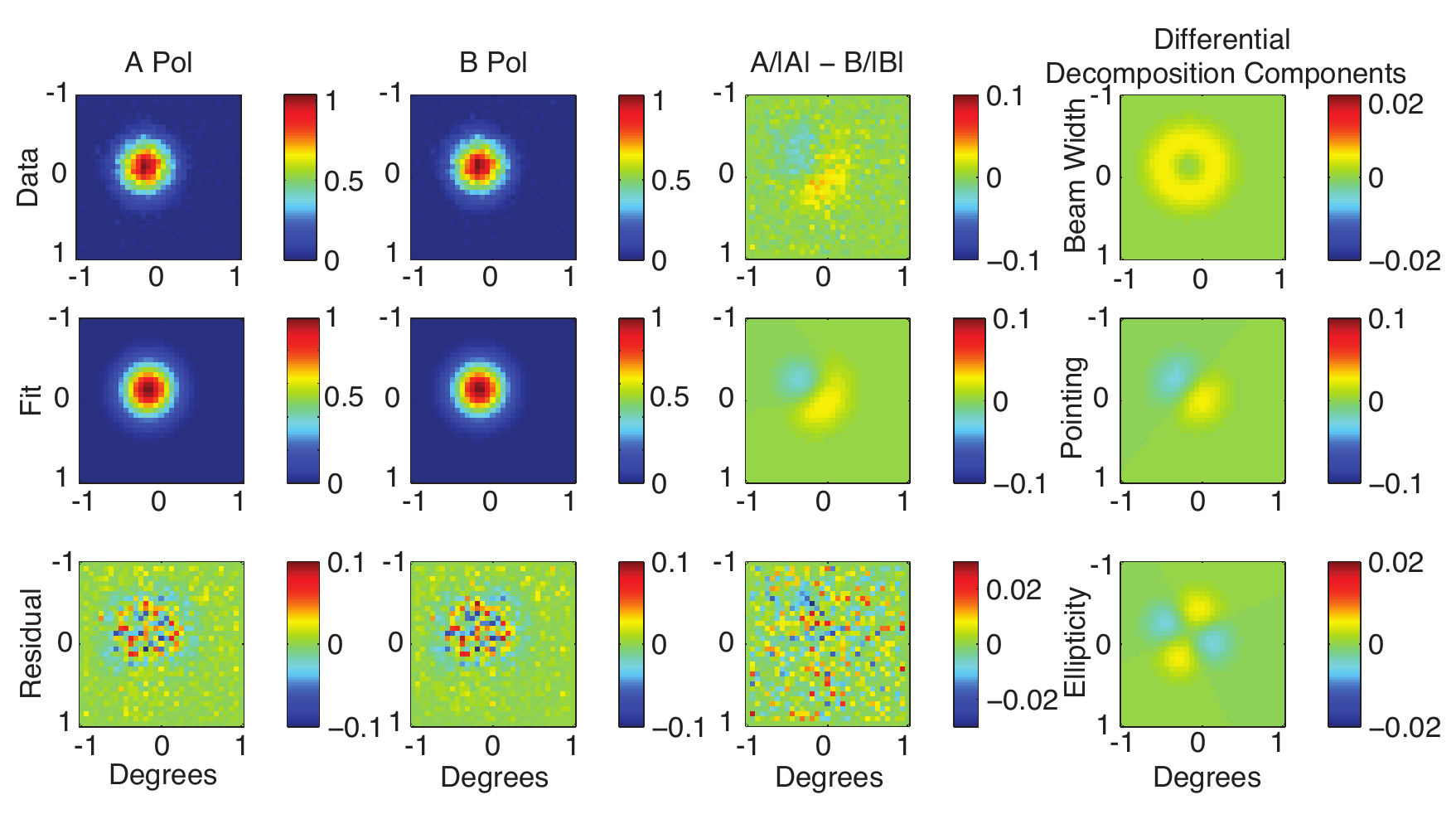}
   \end{tabular}
   \end{center}
   \caption{ \label{fig:multiPanel}
     Example plot of the linear basis fit parameter calculation for one pair of detectors. 
     The first column is the A beam pattern (normalized), the second column is the B beam pattern 
     (normalized), the third column is A-B; 
     for these three columns, the first row is the
     measured map, the second row is the fit, and the third row is the residual. The fourth column is the
     different decomposition components (the template multiplied by the regression coefficient).
}
   \end{figure}

Figure~\ref{fig:multiPanel} shows an example of 
this decomposition for a representative detector pair.  The first column shows the normalized A beam map, 
the second column shows the 
normalized B beam map, and the third column shows the difference between the normalized A and B maps.  We construct 
a model, shown in the second row,
by summing the position offset, beam width, and ellipticity components calculated via template regression for each
detector.  The residual after subtracting the model is shown in the third row.  The fourth column explicitly shows 
the differential components (templates scaled by the regression coefficient) that are summed to give the 
model for the normalized A-B difference map.  The normalization explicitly sets the responsivity mismatch to zero.

\begin{figure}[ht]
   \begin{center}
   \begin{tabular}{c}
   \includegraphics[height=3.3cm]{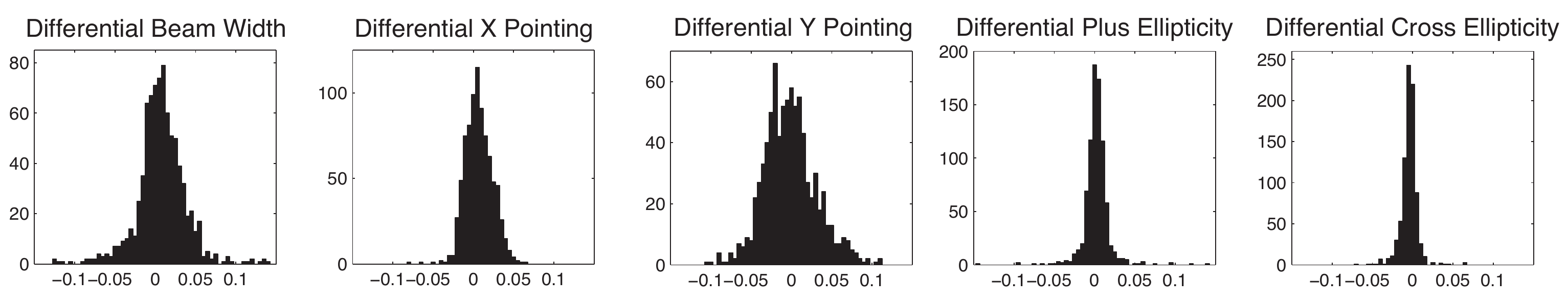}
   \end{tabular}
   \end{center}
   \caption{ \label{fig:beamHistos}
   Histograms of differential beam parameters.  Regression coefficients for each of the templates shown 
   in Figure~\ref{fig:6params} are calculated individually for A and B 
   detectors in a pair and are then differenced to extract differential beam parameters.
}
   \end{figure}

Histograms of all differential 
parameters for each detector pair 
in the Keck Array are shown in Figure~\ref{fig:beamHistos} and receiver-averaged values are shown
in Table~\ref{table:diffParamNumbers}.  Each measurement of each 
detector's main beam must pass a set of criteria to be included in the final extraction 
of beam parameters, including
a check that the beam center was not near the edge of the mirror and excluding measurements
where the initial elliptical Gaussian fit failed.
Note that the relative amplitudes of the distinct
differential parameters shown in the histograms and in Table~\ref{table:diffParamNumbers} 
do not reflect the relative map amplitudes (templates scaled by the regression coefficient).
\begin{table*}[h]
\small
\begin{center}
\begin{tabular}[c]{|l|c|c|c|c|c|}
\hline
Parameter               &   Receiver 0 & Receiver 1 & Receiver 2 & Receiver 3 & Receiver 4    \\
\hline\hline
Differential beam width ($|\delta \sigma|$)   & $0.014\pm0.015$ & $0.024\pm0.024$ & $0.025\pm0.035$ & $0.023\pm0.020$ & $0.031\pm0.028$ \\
Differential pointing ($\sqrt{\delta r_x^2 + \delta r_y^2}$)	& $0.033\pm0.018$ & $0.037\pm0.016$ & $0.019\pm0.014$ & $0.031\pm0.024$ & $0.035\pm0.024$ \\
Differential ellipticity ($\sqrt{\delta p^2 + \delta c^2}$)   & $0.011\pm0.007$ & $0.013\pm0.008$ & $0.011\pm0.012$ & $0.013\pm0.011$ & $0.015\pm0.015$\\
\hline
\end{tabular}
\end{center}
\caption{Receiver-averaged differential beam parameters with the standard deviation of these 
  parameters across each receiver.  
  Regression coefficients for each of the templates shown 
  in Figure~\ref{fig:6params} are calculated individually for A and B 
  detectors in a pair and are then differenced to extract differential beam parameters.
  Both real scatter among detector pairs and measurement uncertainty for individual detector pairs contribute
  to the spread, but measurement uncertainty 
  is subdominant, especially for the differential pointing parameter}
\label{table:diffParamNumbers}
\end{table*} 

This set of measurements of single normalized A-B difference residuals 
is noise-dominated (see Figure~\ref{fig:multiPanel}).
Increased signal-to-noise beam map data would help improve the data quality.
We will have the opportunity to repeat this beam mapping campaign during the upcoming 
2012-2013~deployment season, but with a higher signal level chopped thermal source
and an amplified circularly polarized broadband noise source.


\section{MODELED DIFFERENTIAL BEAM EFFECTS}
\label{sec:differential}
The largest of the clearly modeled differential beam effects is the pointing mismatch, 
indicating that the Keck Array has a pointing offset between orthogonally-polarized detector pairs
that averages to $1.6-3.1$\% of the beam full width at half maximum (FWHM) 
for different receivers.  We have spent a
great deal of effort characterizing the mismatch and investigating its possible sources. A similar level of 
mismatch is observed in \bicepp~\cite{randol}, but was not present in \bicep~\cite{takahashi}, which 
used horn antennas rather than planar antenna arrays.
\bicep had a very similar optical design to \bicepp and the Keck Array, using the same materials for lenses
and filters.

We have measured the pointing offset extensively in both the near field and the far field, in testbed locations
in North America as well as in situ at the South Pole.
We have seen some evidence that the amplitude of the far-field offset scales with the 
near-field offset.  However, there appears to be a complex relationship between
the overall observed near-field mismatch and the overall far-field mismatch.  


Since deploying the Keck Array to the South Pole, 
we have made great progress in reducing the size of the near-field mismatch arising at the focal plane.  This 
effort is described in detail in O'Brient {\it et al.}~\cite{obrientSPIE}, a companion paper at this conference.  
Recent
measurements have revealed that newly-produced replacement Keck Array detector tiles exhibit a near-field mismatch
that is a factor of 10 smaller than in focal planes currently installed at the South Pole.  

Upcoming detailed measurements of the new reduced near-field mismatch focal planes in the far field 
will be very informative.  We expect a reduction of the far-field mismatch, but because of the apparent complex
relationship between the near and far-field effects, we do not necessarily expect a linear scaling between
the measured reduction of the near-field mismatch and the far-field mismatch.  



\section{RESIDUAL DIFFERENTIAL BEAM EFFECTS}
Temperature to polarization leakage in CMB maps that arises from the linearly-modeled 
components of the differential beams described in the previous section can be mitigated in 
analysis.  A discussion of the methods we are implementing to reject these linear components in 
CMB mapmaking, which does not rely on measurement of their amplitude, can be found in 
Aikin {\it et al.}~\cite{methods}. 
After this mitigation, temperature to polarization leakage due to any unmodeled residual differential 
beam shape will remain.

\begin{figure}[ht]
   \begin{center}
   \begin{tabular}{c}
   \includegraphics[height=8cm]{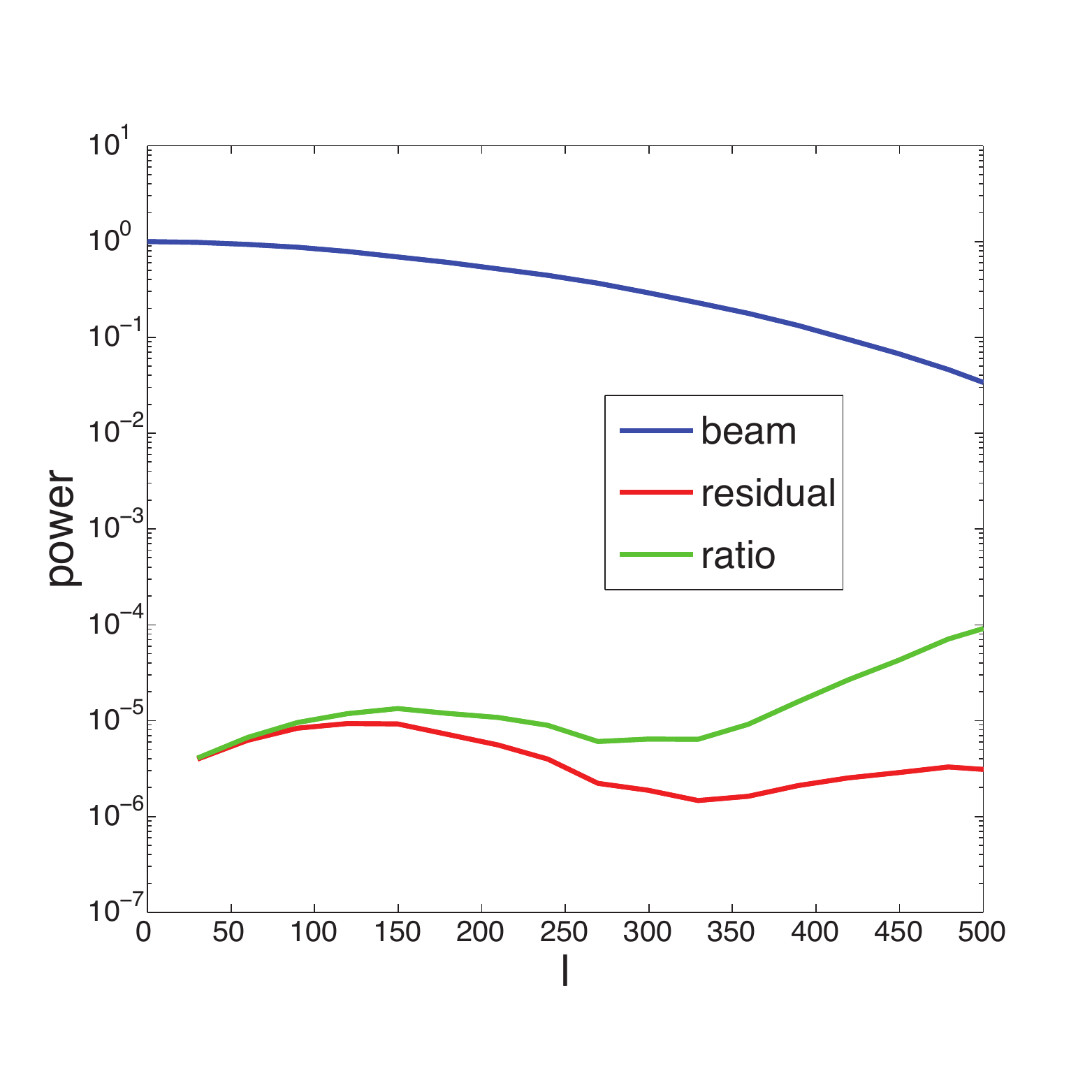}
   \end{tabular}
   \end{center}
   \caption{ \label{fig:leakage}
     Power vs. angular scale for a typical example beam pattern (blue) and the 
     residual beam component after removing leading-order differential components (red).  
     The maps are shown in Figure~\ref{fig:multiPanel}.  The ratio of the 
     two is shown in green.
}
   \end{figure}

Figure~\ref{fig:leakage} shows power vs. angular scale for an example measured beam pattern 
and for the differential
A-B residual after removing the extracted differential components. The A-B residual describes the 
higher-order differential beam effects that remain after mitigation of leading-order effects. To 
reduce the noise floor of the measurement of this residual, we have averaged over all five measurements 
of these detectors after accounting for drum orientation rotation. Still, because of the limited 
signal-to-noise of these measurements, the residual power shown here is only an upper limit on the 
actual residual differential power. We intend to improve the beam map measurements next season to 
achieve higher signal-to-noise measurements by using brighter sources.

At $\ell=100$, where our sensitivity to the Inflationary B-mode spectrum is best, the ratio 
between the unmodeled residual differential power that will remain after mitigation of 
leading-order effects in analysis and the power in the main beam is at most at the $10^{-5}$ 
level for {\it one} typical pair of detectors only.  This upper limit is already close to the raw 
rejection ratio needed to probe polarization to a level of $r=0.01$.  In actual CMB observation, 
we also benefit enormously from averaging-down effects from observing the sky with many detectors 
and at multiple drum angles.  Full simulations of these effects are planned, and together with 
improved beam measurements to constrain residual differential power to lower levels, the 
situation looks promising for control of beam systematics to below the $r=0.01$ level.


\section{CONCLUSIONS}
\label{sec:conclusions}
Pushing deeper into the level of B-mode polarization to constrain $r$ 
requires the dramatically increased sensitivity that the Keck Array provides, but 
also requires tight control of systematics. Through a massive beam mapping campaign, 
we have measured beam properties of each of the 2480~detectors in the Keck Array. We 
extract differential beam parameters using the same linear basis that will be employed 
for analysis mitigation of these modeled effects.  The source of the dominant pointing 
mismatch is still under investigation.  We believe that there is a complex relationship
between the size and orientation of the near-field mismatch and that of the far field.

Measurements with new reduced near-field mismatch focal planes in the far field are 
upcoming.  We expect that the next generation of Keck Array focal planes will benefit 
from dramatically reduced far-field mismatch.  For the current receivers, we have shown 
that noise-dominated upper limits placed on the unmodeled residual component of the 
difference beam pattern already reach the $10^{-5}$ level at $\ell=100$ when including one pair 
of detectors only.  We are optimistic that improved constraints on this residual, together 
with full simulations which account for averaging-down effects from observing the sky with 
many detectors at multiple drum angles, will demonstrate beam systematic control more than 
sufficient to reach $r=0.01$.


\acknowledgments     

The Keck Array is supported by the National Science Foundation, Grant~No.~ANT-1044978/ANT-1110087, and 
by the Keck Foundation.  AGV gratefully acknowledges support from the National Science Foundation, 
Grant~No.~ANT-1103553.  We are 
also grateful to Robert Schwarz for spending the winter at the South Pole in both 2011 and 
2012 for this project, and to the South Pole Station logistics team.  
We thank our \bicepp, Keck Array, and \spider colleagues for useful discussions and shared expertise.


\bibliography{vieregg_spie}   
\bibliographystyle{spiebib}   

\end{document}